\documentclass[10pt,letterpaper]{article}
\usepackage{authblk,amsmath,amssymb,graphicx,geometry,cite,color,mathptmx,courier,textcomp,verbatim}

\usepackage{cite} 
\usepackage[margin=1.27cm]{caption}

\begin{document}

\title{Mid-infrared supercontinuum generation in tapered chalcogenide fiber for producing octave-spanning frequency comb around 3 $\mu$m}

\author{Alireza Marandi,$^{1,*}$ Charles W. Rudy,$^1$ Victor G. Plotnichenko,$^2$ Evgeny M. Dianov,$^2$ Konstantin L. Vodopyanov,$^1$ and Robert L. Byer$^1$}

\affil{$^1$ E. L. Ginzton Laboratory, Stanford University, CA 94305, USA 

$^2$ Fiber Optics Research Center of the Russian Academy of Sciences, Russian Federation}
\affil{* marandi@stanford.edu}
\maketitle

\begin{abstract}
We demonstrate mid-infrared (mid-IR) supercontinuum generation (SCG) with instantaneous bandwidth from 2.2 to 5 $\mu m$ at 40 dB below the peak, covering the wavelength range desirable for molecular spectroscopy and numerous other applications. The SCG occurs in a tapered As$_2$S$_3$ fiber prepared by \emph{in-situ} tapering and is pumped by femtosecond pulses from the subharmonic of a mode-locked Er-doped fiber laser. Interference with a narrow linewidth c.w. laser verifies that the coherence properties of the near-IR frequency comb have been preserved through these cascaded nonlinear processes. With this approach stable broad mid-IR frequency combs can be derived from commercially available near-IR frequency combs without an extra stabilization mechanism. 
\end{abstract}

\section{Introduction}
Broad mid-IR frequency combs \cite{mid-ir-comb} in the 2-20 $\mu$m wavelength range are desirable for many applications, such as molecular fingerprinting \cite{finger}, trace gas detection \cite{trace-gas}, laser-driven particle acceleration \cite{accel}, and x-ray production via high harmonic generation \cite{xray}; however, they are not as easy to produce as near-IR and visible frequency combs. Subharmonic generation using a degenerate optical parametric oscillator (OPO) has been shown to be an effective way to coherently translate a broad near-IR frequency comb to the mid-IR \cite{opex, cleo11, opex12}. Despite the ability of these OPO's to significantly broaden bandwidth, the quest for multi-octave-spanning frequency combs in mid-IR continues.

SCG sources, since their inception in 1970 \cite{scg}, have been extending from the visible into the IR. As low loss optical fibers became prevalent, SCG sources quickly shifted to guided fibers \cite{dudley}, and as the wavelength increased, the glass host has migrated from silica to mid-IR glasses --mainly the chalcogenides, which include the sulfides, selenides, and tellurides \cite{egchalc}. Due to their low loss in the mid-IR, as well as high nonlinearity, hundreds of times higher than silica, the chalcogenides have been an attractive option for an SCG host material \cite{egchalc}. 

In order to achieve significant broadening, the zero group velocity dispersion (GVD) point of the fiber should be close to the center wavelength of the pump (with the pump in the anomalous GVD regime). This minimizes the walkoff between the different regions of the generated spectrum, allowing for further spectral broadening \cite{dudley}. While the material zero GVD point of most chalcogenides is in the mid-IR, GVD can be tailored for efficient SCG \cite{dudley} through either fiber tapering \cite{eggleton}, microstructured fibers \cite{domachuk, micros-sangh, micros2}, or even both \cite{taper-micros}. 

Most previous mid-IR SCG experiments utilize pump lasers in near-IR and expand the spectrum to the mid-IR \cite{eggleton, domachuk, micros2, micros-sangh, taper-micros, nkt, smfiber}; however, since the required broadening is extremely large, none of these methods achieve low-noise SCG \cite{scgnoise2, scgnoise1} and mid-IR frequency combs. Starting with a mid-IR pump would be ideal to achieve a broad spectrum throughout the mid-IR, while operating in the low-noise SCG regime. 

In this paper, we use a subharmonic OPO to convert the ultrashort pulses of a conventional 1.5-$\mu m$ frequency comb source to the mid-IR, while maintaining its coherence properties \cite{cleo11}. After passing the OPO output through a tapered fiber, a mid-IR frequency comb is achieved, centered around 3.1 $\mu m$. This broad source covers the region where many molecules have spectral signatures, and therefore it is ideal for several applications, including dual-comb spectroscopy \cite{keilmann, dual-comb}, and trace-gas detection \cite{trace-gas}.

\section{Design considerations}

\begin{figure}[tbp]
 \centering
 \includegraphics[width=12cm]{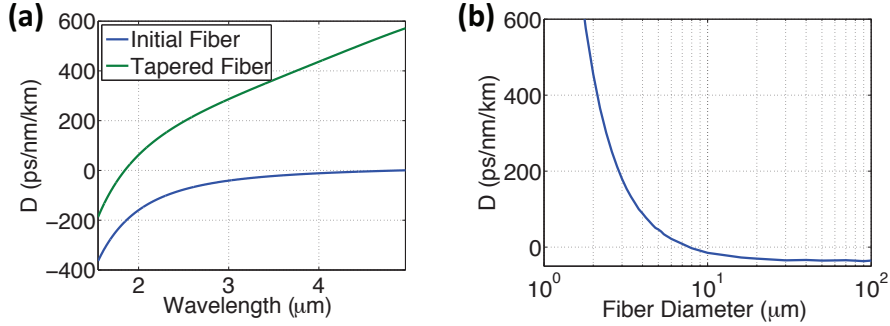}
\caption{ (a) GVD parameter of the initial fiber, a step index As$_2$S$_3$ fiber with core and cladding diameter of 7 $\mu$m and 160 $\mu$m, respectively, and an NA of 0.2, compared with that of the fundamental mode of a tapered fiber with diameter of 2.3 $\mu$m, (b) GVD parameter of the fundamental mode of the tapered fiber at 3.1 $\mu$m as a function of the fiber diameter. }
\label{fig:disp}
\end{figure}

A step index As$_2$S$_3$ fiber is chosen for supercontinuum generation in mid-IR because of its large nonlinear index $\simeq 6 \times 10^{-18}$ m$^2$/W (220 times of that of silica in near-IR) \cite{n2}, and low loss in mid-IR (less than 100 dB/km at 3.1$\mu$m) \cite{dianov}. As shown in Fig. \ref{fig:disp}(a) \cite{sellm}, the zero GVD wavelength of this material is about 4.8 $\mu$m resulting in normal GVD for the fiber in the wavelength range of interest. Tapering the fiber to a few micron in diameter can provide large waveguide dispersion for the fundamental mode and shift the total GVD well into the anomalous regime, as depicted in Fig. \ref{fig:disp}(a). The amount of GVD can be tuned by controlling the diameter of the tapered fiber as depicted in Fig. \ref{fig:disp}(b). Therefore, a tapered As$_2$S$_3$ fiber can be optimized for SCG around the subharmonic of the Er-doped fiber laser at 3.1 $\mu$m.

For fiber tapering, we use the static tapering method, in which a stationary heater is used and the fiber is pulled evenly from both sides with a slow constant speed  \cite{birks}. The diameter of the tapered fiber is then given by:
\begin{equation}
d_{taper}=d_{initial}exp(-L_{pull}/2L_{H}),
\end{equation}
where $d_{taper}$ is the tapered fiber diameter, $d_{initial}$ is the outer diameter of the initial fiber, $L_{pull}$ is the pulling length, and $L_{H}$ is the length of heat zone. Even though this method results in a short tapered fiber region and long transition profile \cite{birks} it is simpler and has higher yield compared to dynamic tapering methods \cite{birks, eggleton}. 

One essential consideration is preserving the coherence of the frequency comb by operating in the low-noise SCG regime \cite{hansch}. In a highly nonlinear tapered fiber, nonlinear amplification of input noise can be significant, even with a shot noise limited input \cite{scgnoise1, scgnoise2}. This noise process, arising from the same fundamental nonlinearity in the fiber, can broaden the linewidth of each comb line and ultimately destroy the comb coherence. More efficient spectral broadening is proven to result in lower noise amplification \cite{scgnoise1,scgnoise2}. Moreover, as the pulses propagate along the fiber, when the rate of spectral broadening starts to decrease, the rate of coherence degradation increases \cite{scgnoise2}. The mentioned observations suggest that optimizing the GVD for spectral broadening and decreasing the length of the tapered fiber can significantly improve the noise in the SCG, which would preserve the coherence properties of the frequency comb.

However, achieving the optimum GVD in the tapered fiber is challenging. When estimating the optimum GVD, the peak power in the tapered fiber must be known \cite{dudley}. It depends on the GVD in the transition region of the fiber and the coupling efficiency, which are not trivial to estimate and could change from one experiment to another. Furthermore, the strong dependence of the GVD on the fiber diameter as shown in Fig. \ref{fig:disp}(b), would make the SCG efficiency very sensitive to the repeatability of the tapering process. Our approach to overcome these challenges and achieve close-to-optimum spectral broadening is \emph{in-situ} tapering. We couple the mid-IR OPO output to the fiber and measure the spectral broadening during the tapering process and use it as the criteria to stop the tapering process.

\section{Experimental setup}

\begin{figure}[tbp]
 \centering
 \includegraphics[width=13cm]{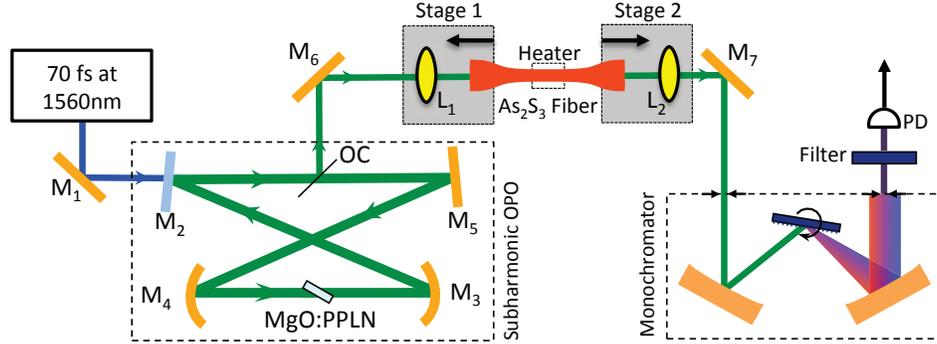}
\caption{Schematic of the experimental setup. }
\label{fig:sch}
\end{figure}

Figure \ref{fig:sch}(a) shows the experimental setup comprised of a commercial mode-locked Er-doped fiber laser, a subharmonic OPO, a tapering apparatus for the As$_2$S$_3$ fiber (with a heat zone length of $L_H=$ 2.1 mm and a heater temperature of $\sim$200 $^\circ$C), and a monochromator. The OPO is synchronously pumped by a 1560-nm mode-locked Er-fiber laser (Menlo Systems C-fiber, 100 MHz, 70 fs, 300 mW) at degeneracy, and provides 47 mW of mid-IR output with pulses shorter than 100 fs centered around 3.1 $\mu$m. The OPO is intrinsically phase and frequency locked to the pump. Details of the OPO operation and its coherence properties are reported in \cite{opex, cleo11}. 

During the tapering process, the output of the OPO is coupled to the fundamental mode of the fiber using a 12.7-mm focal length aspheric ZnSe lens. The fiber is housed in a static tapering setup \cite{birks}, which is based on resistive heating and subsequent pulling of a fiber with core and cladding diameter of 7 $\mu m$ and 160 $\mu m$, respectively, and an NA of 0.2. Stages 1 and 2 are motorized and pull the fiber symmetrically at a constant speed of 10 $\mu$m/s each. The tapered fiber output is collimated with a 20-mm focal length ZnSe plano-convex lens and sent to a monochromator. The monochromator is set to the longer edge of the initial spectrum, around 3.9 $\mu$m with a 20-nm bandwidth, where the photodetector output is a few dB above noise level. This signal is used as a measure of the spectral broadening.

\section{Results and discussions}

\begin{figure}[tbp]
 \centering
 \includegraphics[width=8cm]{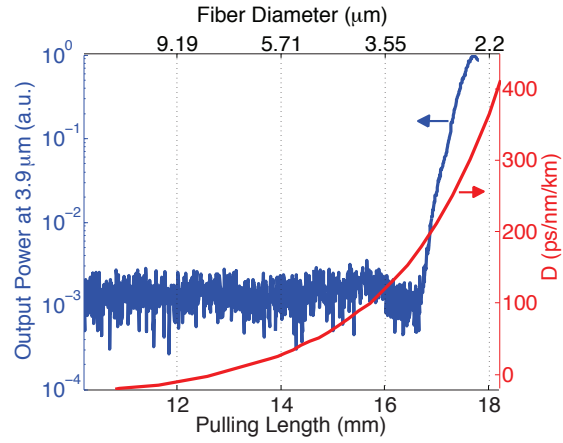}
\caption{The output power at 3.9 $\mu$m during the tapering process (blue curve) versus the pulling length (bottom axis) and corresponding diameter of the fiber waist (top axis), and the calculated GVD parameter of the fundamental mode in the taper waist (red curve). }
\label{fig:pulling}
\end{figure}

Figure \ref{fig:pulling} shows the photodetector signal during the tapering process as a function of the pulling length, or the distance added to the initial spacing between stages 1 and 2. Rapid growth of this signal is observed around pulling length of 16.7 mm and reached a maximum at 17.72 mm; slightly afterwards, the tapering process is stopped at 17.82 mm, corresponding to a taper diameter of 2.3 $\mu$m. The behavior of this signal verifies the significant sensitivity of the SCG to the pulling length and taper diameter. For example, from Fig. \ref{fig:pulling}, a taper diameter variation of 126 nm results in 3 dB decrease of the output power at 3.9 $\mu$m, and a 286 nm deviation results in 10 dB decrease. This sensitivity is associated with the strong dependence of the GVD of the fiber on its diameter, and the amount of GVD in the transition region, as discussed earlier.
In several successful tapering experiments the deviation in the pulling length's peak position was around 0.3 mm, corresponding to difference in either the repeatability of tapering or the amount of GVD before the tapered fiber. These observations emphasize the necessity of \emph{in-situ} tapering.


\begin{figure}[tbp]
 \centering
 \includegraphics[width=12cm]{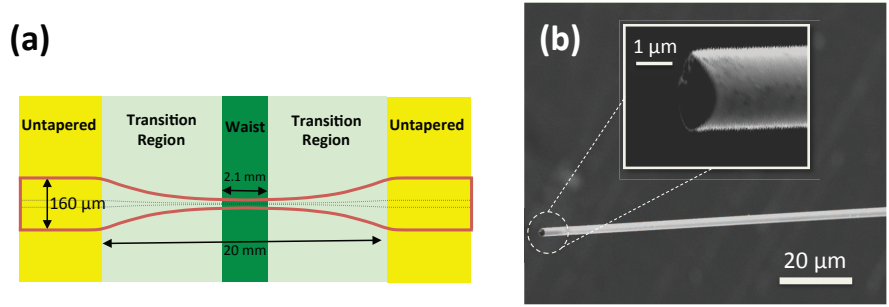}
\caption{(a) Schematic profile of the tapered fiber; the length of the untapered fiber is $\sim$18 mm on each side; the schematic is not to scale. (b) SEM images of the tapered fiber broken at the waist. }
\label{fig:sem}
\end{figure}

The resulting tapered fiber has a waist length of $\sim$ 2.1 mm, the same as the heat length, with a profile as illustrated in Fig. \ref{fig:sem}(a). The exponential profile of the transition region has been observed by scanning electron microscope (SEM) imaging of different sections after the experiment. The SEM image of one of the successfully tapered fibers broken after the experiment is shown in Fig. \ref{fig:sem}(b). The waist diameter is 2.3 $ \mu$m, corresponding to a nonlinearity of $\gamma\simeq$ 4.5 W$^{-1}$ m $^{-1}$ \cite{dudley}. We could remove the tapered fiber from the tapering setup several times without breaking it, but we did not perform any optical characterization outside the setup. The mechanical strength of such tapered fibers can be significantly increased by having an additional cladding of appropriate material similar to \cite{pmma_clad}.

The mid-IR output power after the collimating lens is measured to be 15 mW, corresponding to the pulse energy of 150 pJ in free space and 250 pJ in the fiber. Before the tapering process, the power out of the fiber is 16 mW. This difference is mostly due to slight coupling to the higher-order or cladding modes, for which the tapered fiber has a large loss. The loss associated with the tapering, from the untapered fiber at the input to the untapered fiber at the output is lower than 6\%, which is also the measurement accuracy of the detector. A pyroelectric camera has been used to image the face of the fiber at the output to distinguish between the mid-IR transmission through the fundamental and cladding modes of the fiber. Before tapering, the coupling to the fundamental mode is optimized, and after tapering it is verified that the mid-IR light is still in the fundamental mode at the output.

\begin{figure}[tbp]
 \centering
 \includegraphics[width=8cm]{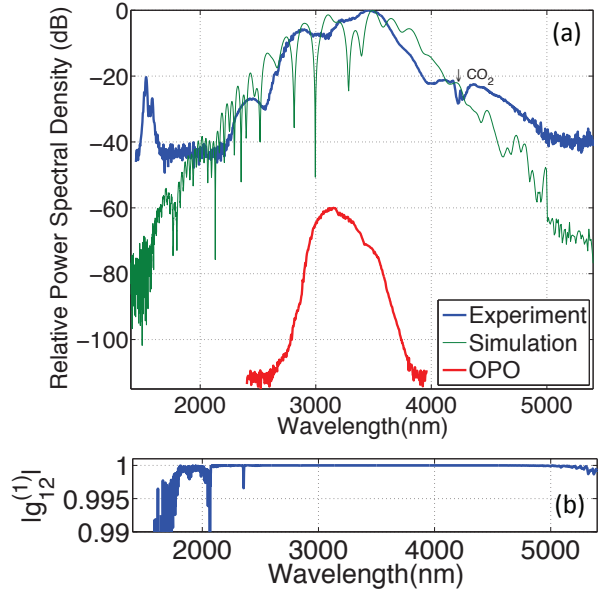}
\caption{(a) Spectrum of the OPO output broadened by the tapered As$_2$S$_3$ fiber compared with the simulation result (the OPO spectrum is intentionally shifted) , (b) calculated degree of coherence, $g^{(1)}_{12}$  as defined in \cite{scgnoise2}. }
\label{fig:spectrum}
\end{figure}

The spectrum is measured by an $f$ = 20 cm monochromator equipped with a 100 line/mm diffraction grating and a thermoelectrically cooled InSb detector, with a resolution of about 20 nm. To block any second order diffraction of the grating overlapping with the spectrum, an InAs filter was used to measure the long wavelength portion of the spectrum ($>$ 4.1 $\mu$m).
The spectrum of the fiber output is depicted in Fig. \ref{fig:spectrum}(a) extending from 2.2 $\mu m$ to 5 $\mu m$ at 40 dB below the peak, and is about three times broader in frequency bandwidth than the OPO output spectrum as shown on the same plot. The output spectrum is extended from 2.3 to 4.7 $\mu$m at 30 dB below the peak, and 2.8 to 3.7 $\mu$m at 10 dB below the peak. The CO$_2$ absorption signature is observable around 4.2 $\mu m$ due to passing through the atmosphere between the tapered fiber and the detector. The signal around 1.6 $\mu$m is due to the transmission of the fiber laser output through the tapered fiber.



The nature of the SCG process can be perceived by estimating the nonlinear parameters \cite{dudley} of this experiment. The nonlinear length for the tapered fiber is $L_{NL}\simeq$ 0.15 mm, and the dispersion length is $L_D \simeq$ 9.8 mm, resulting in a soliton order of $N\simeq \sqrt{L_{D}/L_{NL}}\simeq$ 8 and a fission length of $L_{fiss}\simeq$ 1.2 mm. These values suggest that the spectral broadening is dominated by soliton fission \cite{dudley}. Based on previous studies of SCG in the near-IR \cite{genty}, these parameters suggest a high degree of coherence with a large margin before coherence degradation. Hence, the spectral broadening can be further improved by either increasing the peak power or the taper length, and a high degree of coherence and frequency comb at the output would still be achieved.

Numerical simulation of generalized nonlinear Schrodinger equations \cite{dudley} has been used and the resulting spectrum is in good agreement with the experimental result as shown in Fig. \ref{fig:spectrum}(a). The simulation is performed for the 2.1 mm length of tapered As$_2$S$_3$ fiber, using the Raman fraction and response reported in \cite{raman_coef}. For the input pulse, the OPO output is assumed to be transform-limited and has propagated in the untapered fiber and transition region with the calculated average GVD of -19 ps/nm/km. Slight differences between the simulation and experiment can be due to broadening in the transition region, as well as a non-uniform spectral response in the experimental setup, which is largely unaccounted for in the simulation. The oscillations in the simulated spectrum are due to the pulse breaking, which were not captured in the measured spectrum because of the low spectral resolution. The degree of coherence, defined in \cite{scgnoise2} as  $g^{(1)}_{12}$, is also calculated using the same simulation tool and is depicted in Fig. \ref{fig:spectrum}(b). This value is larger than 0.995 for the whole range of wavelengths above the noise level in the experiment, which confirms operation in the low-noise SCG regime, and therefore a coherent output from 2.2 to 5 $\mu$m.

\begin{figure}[tbp]
 \centering
 \includegraphics[width=8cm]{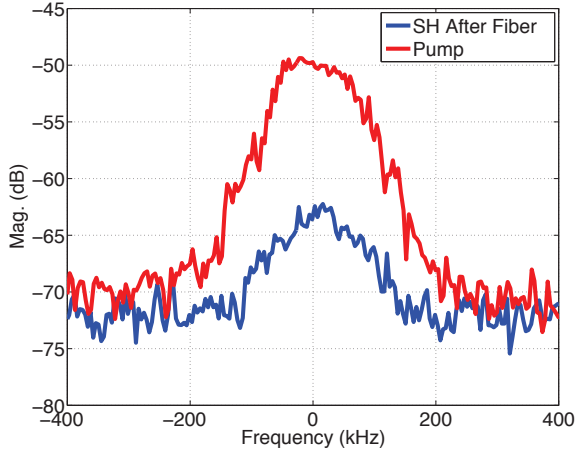}
\caption{RF spectrum of the interference of a c.w. laser at 1564 nm with the free-running Er-doped fiber laser and the second harmonic (SH) of the tapered fiber output. }
\label{fig:BN}
\end{figure} 

To verify that the broad mid-IR output has a frequency comb structure and examine its characteristics, a c.w. laser at wavelength of 1564 nm with 3-kHz linewidth is used. The mid-IR output is first filtered using a Ge filter to ensure no residual signal at 1.56 $\mu$m exists, and then it is focused into a 1-mm long periodically-poled lithium niobate crystal to generate the second harmonic. The second harmonic is then interfered with the c.w. laser using a fiber coupler. The interference signal is measured by a fast InGaAs detector and is compared with the interference signal of the mode-locked fiber laser and the c.w. laser. The beat frequencies in both cases are the same, and they track each other during tuning of the carrier-envelope frequency of the mode-locked laser. The RF spectrum of these signals are depicted in Fig. \ref{fig:BN}. They have the same full-width at half-maximum of 140  KHz which is the comb linewidth of the free-running mode-locked laser. This experiment and simulation results verify that the coherence properties of the initial frequency comb in the near-IR are preserved and are intrinsically transferred to the mid-IR comb, through the subharmonic generation by the OPO and the subsequent spectral broadening in the tapered fiber.

\section{Conclusion}

We demonstrate SCG in a tapered chalcogenide fiber that spans more than an octave in mid-IR, which is ideal for molecular spectroscopic applications. We show how the coherence properties and the spectral comb structure of a commercially available near-IR frequency comb can be transferred to mid-IR through the cascaded subharmonic and supercontinuum generations. Numerical simulations suggest that further broadening of the spectrum, while maintaining high coherence, is possible. This can be achieved by increasing either the length of the tapered fiber or the pump peak power.  

\section*{Acknowledgements}
The authors would like to thank G. Shambat, C. Phillips, K. Aghaei for invaluable discussions, F. Afshinmanesh for SEM images, T. Marvdashti for experimental support, and M. F. Churbanov and G. E. Snopatin from the Institute of Chemistry of High-Purity Substances for providing the As$_2$S$_3$ fiber.

\end{document}